\documentclass[aps,prb,superscriptaddress,showpacs,amsmath,twocolumn]{revtex4}

\usepackage{amsmath}
\usepackage{amssymb}
\usepackage{graphicx}

\newcommand{\pfi}{p_{\varphi}}
\newcommand{\prt}{\partial}

\newcommand{\vfi}{\varphi}

\begin{document}

\title{Bound states in inhomogeneous magnetic field in graphene: a semiclassical
 approach}

\author{A.~Korm\'anyos}
\affiliation{Department of Physics, Lancaster University,
Lancaster, LA1 4YB, UK}

\author{ P. Rakyta
%\thanks{e-mail: cserti@complex.elte.hu}
}
\affiliation{Department of Physics of Complex Systems,
E{\"o}tv{\"o}s University,
H-1117 Budapest, P\'azm\'any P{\'e}ter s{\'e}t\'any 1/A, Hungary
}

\author{L.~Oroszl\'any}
\affiliation{Department of Physics, Lancaster University,
Lancaster, LA1 4YB, UK}

\author{ J. Cserti
\thanks{e-mail: cserti@complex.elte.hu}
}
\affiliation{Department of Physics of Complex Systems,
E{\"o}tv{\"o}s University,
H-1117 Budapest, P\'azm\'any P{\'e}ter s{\'e}t\'any 1/A, Hungary
}

%\wideabs{

\begin{abstract}
We derive semiclassical quantization equations for graphene mono- and 
bilayer systems where  the excitations are confined by the  
applied inhomogeneous magnetic field.  The importance of a semiclassical 
phase, a consequence of the spinor nature of the excitations, 
is pointed out. The semiclassical eigenenergies show
good agreement with the results of  quantum mechanical calculations based on the 
Dirac equation of graphene and with  numerical tight binding calculations.

\end{abstract}


\maketitle

\section{Introduction}

Observation of  massless Dirac fermion type excitations 
in recent experiments on graphene has  generated
 huge interest both experimentally and
theoretically~\cite{Novoselov-1:cikk,Kim:cikk,Novoselov-bilayer:cikk}.
For reviews on graphene see
Refs.~\onlinecite{Katsnelson:rev,Katsnelson_Novoselov:rev,Geim_Novoselov:rev,
ref:guinea-1,ref:guinea-2}
and a special issue in Ref.~\onlinecite{Solid_State_Comm:cikkek}.

The intensive theoretical and experimental work have lead to good  
understanding of the physical phenomena in the bulk of disorder-free  
graphene in homogeneous magnetic field\cite{ref:dirac_LL,ref:falko}. 
Recently, the interest in inhomogeneous magnetic field setups has 
also appeared. Martino et al.~\cite{ref:martino} have
demonstrated that  massless Dirac electrons can be confined by
inhomogeneous magnetic field and that a magnetic
quantum dot can be formed in graphene, levels of which are  
tunable almost at will. The so-called "snake 
states" known from studies\cite{Muller:cikk,non-uniB:cikkek} 
on two dimensional electron gas (2DEG) have  also attracted 
interest and their properties have been discussed in graphene 
monolayer\cite{ref:rakyta,ref:ghosh} and 
in carbon-nanotubes\cite{Nemec:cikk,lee:cikk}.  

Semiclassical methods have helped our understanding  
of complicated physical phenomena enormously and become a standard
tool of investigation. Not only they  offer 
a simple, easy to grasp classical picture but in many cases they can 
also give quantitative predictions on observables. Yet the first 
semiclassical study on graphene systems has only very recently 
appeared\cite{ref:ullmo}. Ullmo and Carmier derived  
and expression for the semiclassical Green's function in graphene and
studied the  "Berry-like"  phase which appears in the
semiclassical theory. The importance of "Berry-like" and 
"non-Berry-like" phases in the 
asymptotic theory of coupled partial differential equations and 
their r\^ole in semiclassical quantization 
 have previously been discussed in 
Refs.~\onlinecite{ref:yabana,ref:littlejohn,ref:gyorffy,ref:keppeler}.

\begin{figure}[hbt]
\includegraphics[scale=0.35]{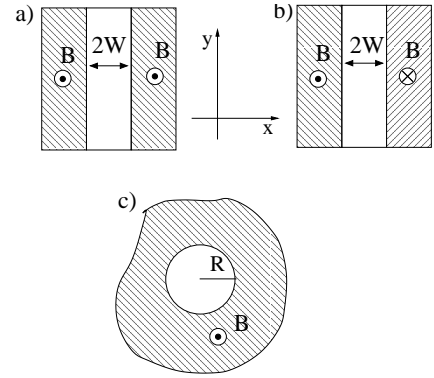}
\caption{
In a) and b) the  applied perpendicular magnetic field 
$\mathbf{B}$   is zero in the center region of  width $2W$.  
In the case of a), in the left and right 
regions the magnetic fields point in the same directions, while in the case
of  b), in the opposite directions. The magnitude $B$ of the magnetic field
is the same in both left and right regions.  
In the case of Fig.~\ref{fig:geometries}(c) 
a circular  non-magnetic region of radius $R$ 
is considered in graphene monolayer, 
whereas outside this region there is perpendicularly applied  
magnetic field of magnitude $B$.  
\label{fig:geometries}}
\end{figure}

In this work we study a graphene nanoribbon\cite{ref:wakabayashi} 
in a non-uniform magnetic 
field\cite{ref:ghosh,ref:martino,ref:rakyta}  as shown in 
Fig.~\ref{fig:geometries}(a) and (b) and a circular magnetic dot 
in graphene monolayer\cite{ref:martino} [see Fig.~\ref{fig:geometries}(c)].
We assume   that the applied  perpendicular magnetic field
of magnitude $|B_z|$ changes in step-function like manner at the 
interfaces of the magnetic and non-magnetic regions and  
that it is strong enough so that the magnetic length  
$l_B=\sqrt{\hbar/e|B_z|}$ is much smaller than the characteristic 
spatial dimension of the graphene sample.
We show that in this case the semiclassical 
quantization can predict and help to understand  the main features of the 
the quantum spectra at the $\mathbf{K}$ point\cite{Katsnelson:rev,Katsnelson_Novoselov:rev,Geim_Novoselov:rev,
ref:guinea-1,ref:guinea-2} of  
graphene mono- and bilayer.

The article is organized in the following way. Firstly, in section \ref{sec:qm} we 
give a  brief  overview of the exact quantum mechanical treatment 
of graphene mono and bilayer. We also discuss  some of the technical details 
of the quantum calculation regarding the system shown in Fig.\ref{fig:geometries}.
Next, in section \ref{sec:sc_formalism} we 
introduce our  semiclassical formalism and  wherever possible
we give a unified description for graphene mono and bilayer. 
In section  \ref{sec:landaugauge} we present the results of the semiclassical
quantization for graphene nanoribbons in inhomogeneous magnetic field and
compare it to tight binding (TB) and exact quantum calculations.  
In the next,  section \ref{sec:symmgauge} we apply the semiclassical
formalism to a magnetic dot in graphene monolayer. Finally, in section
\ref{sec:concl} we arrive to our conclusions.

\section{Monolayer and bilayer graphene: quantum mechanical treatment}
\label{sec:qm}

In the simplest approximation the  Dirac Hamiltonian describing the
 low-energy excitations at the ${\bf K}$ point of the Brillouin zone 
in monolayer (bilayer) graphene reads
\begin{equation}
\hat{H}_{\beta}=g_{\beta}
\left(
\begin{array}{cc}
0 & (\hat{\pi}_x-i\hat{\pi}_y)^{\beta}\\
  (\hat{\pi_x}+i\hat{\pi}_y)^{\beta} & 0 
\end{array}
\right).
\label{eq:qm_hamilt}
\end{equation}
Here $\beta=1\, (2)$ for monolayer (bilayer) and  
 $g_{1}=v_F=\sqrt{3}/2 \, a\, t_1 /\hbar$ is given by the hopping parameter
$t_1$ between
the nearest neighbours in monolayer graphene ($a=0.246$ nm is the lattice
constant in the honeycomb lattice). Moreover, $g_{2}=-1/2 m^*$ and 
the mass term is given by $m^*=t_2/ 2 (g_1)^2$, 
$t_2$ being the interlayer hopping
between  $\tilde{A}-B$ sites of bilayer\cite{ref:falko}. The operators 
$(\hat{\pi}_x,\hat{\pi}_y)$ are defined by 
$ \mbox{\boldmath $\hat{\pi}$}= (\hat{\pi}_x,\hat{\pi}_y)=\hat{\bf p} + e {\bf A}$, 
where  $\hat{\bf p}=\frac{\hbar}{i}\frac{\prt}{\prt \mathbf{r}}$ 
is the canonical momentum operator and the vector potential ${\bf A}$  
is related to the magnetic field through  ${\bf B}= \text{rot}{\bf A}$.
Due to the chiral symmetry $\sigma_z\hat{H}_{\beta}\sigma_z=-\hat{H}_{\beta}$, where
$\sigma_z$ is a Pauli matrix, it is enough to consider the positive eigenvalues 
of the Hamiltonian $\hat{H}_{\beta}$.

In section \ref{sec:landaugauge} we study a 
graphene nanoribbon\cite{ref:wakabayashi}
of width $L\gg l_B$ %$l_B=\sqrt{\hbar/e|B|}$ being the magnetic length 
[see  Figs.~\ref{fig:geometries}(a),(b)]. In the central region 
$|x|< W$ the magnetic field is zero, while for $|x| \ge W$ a non-zero 
perpendicular magnetic field is applied. We assume a step function-like 
change of the magnetic field at $x=W$ and use  the vector potential   $\mathbf{A}(\mathbf{r})=(0,A_y(x),0)^{T}$  to preserve 
the translation invariance in the $y$ direction. 
Other  details of the quantum calculation can be found
in  Refs.~\onlinecite{ref:martino,ref:rakyta,ref:ghosh}.

The magnetic dot system is shown in Fig.~\ref{fig:geometries}(c). 
It consists of a graphene monolayer in homogeneous perpendicular magnetic 
field with a circular inclosure where the magnetic field is zero. The 
circular symmetry of the setup suggests that one should choose the 
vector potential in the symmetric gauge:
\begin{equation}
\mathbf{A}(\mathbf{r})=\left\{
\begin{array}{cc}
0 & r<R, \\
\frac{B_z(r^2-R^2)}{2r}
\left(
\begin{array}{c}
      -\sin\vfi \\
        \cos\vfi\\
       0
\end{array}\right) & r \ge R,
\end{array}
\right.
\label{eq:vect_pot_pol}
\end{equation}
where $\mathbf{r}=(r\cos\vfi,r\sin\vfi)$ is  in polar coordinates.
One can show that with this choice the Schr\"odinger equation 
$
\hat{H}_{\beta}\Psi_{\beta} =E \Psi_{\beta}
$ 
becomes separable in  $r$ and $\vfi$.  
In the case of graphene monolayer, requiring that the
wave function be normalizable and continuous  at $r=R$  
leads to a secular equation, solutions of
which are the quantum eigenenergies (see Eq.~(21) in Ref.~\onlinecite{ref:ghosh}).

\section{Semiclassical formalism for graphene}
\label{sec:sc_formalism}

We now give a brief account of our semiclassical formalism. 
Our discussion goes along the lines of the  Refs.~\onlinecite{ref:gyorffy,ref:keppeler,ref:ullmo} from where we have 
also borrowed some of the notations. 

We seek the solutions of the Schr\"odinger equation 
$
\hat{H}_{\beta}\Psi_{\beta} =E \Psi_{\beta}
$ 
in the following form \cite{ref:keppeler}:
\begin{equation}
\Psi_{\beta}(\mathbf{r})=\sum_{k\ge 0} \left(\frac{\hbar}{i}\right)^{k} 
 \mathbf{a}_k^{\beta}(\mathbf{r}) e^{\frac{i}{\hbar} S_{\beta}(\mathbf{r})},
\label{eq:semiclass_wavef}
\end{equation}
where $\mathbf{a}_k^{\beta}(\mathbf{r})$ are  spinors and 
$ S_{\beta}(\mathbf{r})$ is the classical action. 
Performing the unitary transformation 
$
\Psi_{\beta}\rightarrow e^{-\frac{i}{\hbar} S_{\beta}(\mathbf{r})}\Psi_{\beta}
$, 
$\hat{H}_{\beta}\rightarrow
e^{-\frac{i}{\hbar} S_{\beta}(\mathbf{r})} \hat{H}_{\beta}\,
e^{\frac{i}{\hbar} S_{\beta}(\mathbf{r})}
$
the Schr\"odinger equation can be rewritten  as 
\begin{equation}
\begin{split}
\left(
\begin{array}{cc}
- {E} & g_{\beta}(\hat{\Pi}_x-i \hat{\Pi}_y)^{\beta}\\
  g_{\beta} (\hat{\Pi}_x+i \hat{\Pi}_y)^{\beta} & -{E} 
\end{array}
\right)
\left(\mathbf{a}^{\beta}_0(\mathbf{r})\right.+\\
\left.\frac{\hbar}{i}\mathbf{a}^{\beta}_1(\mathbf{r})
+\dots \right)=0.
\label{eq:semiclass_schr}
\end{split}
\end{equation}
Here 
$\hat{\Pi}_x=\hat{p}_x+ \Pi^{0}_x$, where
$\Pi^{0}_x=p_x +e A_x(\mathbf{r})$, 
$p_x=\frac{\partial S_{\beta}(\mathbf{r})}{\partial x}$, and
similarly for $\hat{\Pi}_y$. The WKB strategy\cite{ref:brack} 
is to satisfy Eq.~(\ref{eq:semiclass_schr}) separately order by order in $\hbar$. 

At $\mathcal{O}(\hbar^0)$ order we obtain 
\begin{equation}
\left(
\begin{array}{cc}
- {E} & g_{\beta}({\Pi}_x^0-i {\Pi}_y^0)^{\beta}\\
  g_{\beta}({\Pi}_x^0+i {\Pi}_y^0)^{\beta} & -{E} 
\end{array}
\right)
\mathbf{a}^{\beta}_0(\mathbf{r})=0.
\label{eq:hbar0ord}
\end{equation}
This classical Hamiltonian can be diagonalized with eigenvalues
\begin{equation}
\mathcal{H}_{\beta}^{\pm}(\mathbf{p},\mathbf{r})=\pm g_{\beta}[(\Pi_x^0(\mathbf{r}))^2+(\Pi_y^0(\mathbf{r}))^2]^{\beta/2}.
\label{eq:class_Ham}
\end{equation}
and eigenvectors $V^{\pm}_{\beta}(\mathbf{p},\mathbf{r})$. 
What we have found is  that the  $\mathcal{O}(\hbar^0)$ order equation is
in fact equivalent to a pair of classical Hamilton-Jacobi equations:
\begin{equation}
E-\mathcal{H}_{\beta}^{\pm}
\left(\frac{\prt S_{\beta}^{\pm}(\mathbf{r})}{\prt \mathbf{r}},\mathbf{r}\right)=0.
\label{eq:HamiltJacob}
\end{equation}
The solution of Eq.~(\ref{eq:HamiltJacob})-when it exists- can be found
 e.g. by the method of characteristics\cite{ref:ullmo}.

For $E\neq 0$ the eigenvectors of the classical Hamiltonian given in 
Eq.~(\ref{eq:hbar0ord}) are 
\begin{equation}
V_{\beta}^{\pm}= 
\frac{1}{\sqrt{2}}
\left(
\begin{array}{c}
\pm (-1)^{\beta-1} e^{-i\beta\phi} \\
1
\end{array}
\right)
\label{eq:V}
\end{equation}
(here $\phi$ is the phase of $\Pi_x^0-i\Pi_y^0$)
but the eigenspinor $\mathbf{a}_0^{\beta, \pm}$ can be more generally written as 
$
\mathbf{a}_0^{\beta, \pm}=\mathcal{A}_{\beta}^{\pm}(\mathbf{r}) e^{i\gamma_{\beta}^{\pm}(\mathbf{r})}V_{\beta}^{\pm}
$
where $\mathcal{A}_{\beta}^{\pm}(\mathbf{r})$ is a real amplitude 
and $\gamma_{\beta}^{\pm}(\mathbf{r})$ 
is a phase.  Equations for $\mathcal{A}_{\beta}^{\pm}(\mathbf{r})$ 
and $\gamma_{\beta}^{\pm}(\mathbf{r})$
can be obtained from the  $\mathcal{O}(\hbar^1)$ order of 
Eq.~(\ref{eq:semiclass_schr}). One can show that 
the $\mathcal{O}(\hbar^1)$ order equation can be written in
the following form\cite{ref:gyorffy,ref:keppeler,ref:ullmo}
\begin{equation} 
(\mathbf{a}_0^{\beta, \pm})^{\dagger}\hat{M}_{\beta}\mathbf{a}_0^{\beta, \pm}=0.
\label{eq:h1_ord}
\end{equation}
Using the notation $\mbox{\boldmath ${\sigma}$}=(\sigma_x,\sigma_y)$, 
$\sigma_{x,y,z}$ being the Pauli matrices, the operator $\hat{M}_1$ 
 for graphene monolayer is 
$\hat{M}_1=\mbox{\boldmath ${\sigma}$}\mathbf{\hat{p}}$, while 
 for bilayer it reads $\hat{M}_2=\hat{m}+\hat{m}^{\dagger}$, where
$\hat{m}=\mbox{\boldmath ${\sigma}$}\mathbf{\hat{p}}(\Pi_x^0+i\sigma_z\Pi_y^0)$.

The imaginary part of  Eq.~(\ref{eq:h1_ord})  expresses the 
conservation of probability 
since it can be cast into the form of  a continuity equation 
$\textnormal{div}\mathbf{j}_{\beta}^{\pm}=0$. Here 
%\begin{equation}
$
\mathbf{j}_{\beta}^{\pm}=\mbox{Im}\langle\Psi_{\beta}^{s,\pm}| 
\mathbf{\hat{v}}_{\beta}|\Psi_{\beta}^{s,\pm}\rangle
$
%\end{equation} 
is the probability current carried by
the semiclassical wavefunction
$\Psi_{\beta}^{s,\pm}=\mathbf{a}_0^{\beta, \pm}e^{\frac{i}{\hbar}S_{\beta}^{\pm}}$
( $\mathbf{\hat{v}}_{\beta}=\frac{i}{\hbar}[\hat{H}_{\beta},\mathbf{r}]$ is 
the velocity operator).
Similarly to the case of quantum systems described by scalar Schr\"odinger 
equation\cite{ref:brack},
this continuity equation determines $\mathcal{A}_{\beta}^{\pm}(\mathbf{r})$.

The real part of  Eq.~(\ref{eq:h1_ord})  allows to calculate the 
phase $\gamma_{\beta}^{\pm}(\mathbf{r})$.  The equation determining $\gamma_{\beta}^{\pm}(\mathbf{r})$ reads
\begin{equation}
\frac{d \gamma_{\beta}^{\pm}(\mathbf{r})}{dt}=c_{\beta}^{\pm}\frac{\beta}{2}
\left(\frac{\prt \Pi_y^0(\mathbf{r})}{\prt x}-
\frac{\prt \Pi_x^0(\mathbf{r})}{\prt y}\right)
=c_{\beta}^{\pm} \frac{\beta}{2} eB_z(\mathbf{r}),
\label{eq:dtgamma}
\end{equation}
where $c_1^{\pm}=v_F^2/E$, $c_2^{\pm}=\pm 1/m^*$, and we denoted by 
$B_z(\mathbf{r})$  the component of the applied magnetic field
which is perpendicular to the graphene sheet.
The second equality in Eq.~(\ref{eq:dtgamma}) follows from 
$
\frac{\prt^2 S_{\beta}(\mathbf{r})}{\prt x \prt y}=
\frac{\prt^2 S_{\beta}(\mathbf{r})}{\prt y \prt x}
$.

The Hamiltonian given in Eq.~(\ref{eq:qm_hamilt}) yields a gapless spectrum 
for bilayer graphene. Theoretical and experimental studies of bilayer 
graphene\cite{ref:falko,ref:ohta} 
have shown  that an electron density dependent gap can exists 
between the otherwise degenerate valence and conductance bends 
(described by $\mathcal{H}_{2}^{-}$ and $\mathcal{H}_{2}^{+}$ respectively,
in our semiclassical formalism). Assuming that the gap is  spatially constant,
one can take it into account by considering the Hamiltonian\cite{ref:falko}
$\hat{H}_{2,\Delta}=\hat{H}_{2}+(\Delta/2)\sigma_z$. The new term 
$(\Delta/2)\sigma_z$  affects only the $\mathcal{O}(\hbar^0)$ calculations,
while the operator $\hat{M}_2$ in Eq.~(\ref{eq:h1_ord}) remains the same.
Consequently, Eq~(\ref{eq:class_Ham}) is modified to 
\begin{equation}
\mathcal{H}_{2}^{\pm}(\mathbf{p},\mathbf{r})=\pm \frac{1}{2m^*}
\sqrt{[(\Pi_x^0)^2+(\Pi_y^0)^2]^2+\tilde{\Delta}^2}.
\label{eq:Ham_bilay_Delta}
\end{equation}
where $\tilde{\Delta}=\Delta/m^*$
and the right hand side of Eq.~(\ref{eq:dtgamma}) for $\beta=2$ is multiplied by 
$\eta=\sqrt{1-\frac{\Delta^2}{4 E^2}}$.

It has been shown in Refs.~\onlinecite{ref:yabana,ref:gyorffy,ref:keppeler}
that for  $N$-dimensional 
integrable systems  where the particles have an internal, e.g. spin 
or electron-hole degree of freedom, one can derive a generalization of the 
EBK\cite{ref:brack}  quantization of scalar systems. In general, the 
quantization conditions read
\begin{equation}
\frac{1}{\hbar}\oint_{\Gamma_j} \mathbf{p}\,d\mathbf{r} + \alpha_j =
2\pi\left(n_j+\frac{\mu_j}{4}\right)
\label{eq:EBK_general}
\end{equation}
Here $\Gamma_j$, $j=1\dots N$  are the irreducible loops on the  
$N$-torus in the phase space,   $n_j$s are  positive integers, 
$\mu_j$s are the Maslov indices\cite{ref:brack} 
counting the number of caustic points along $\Gamma_j$  and finally 
$\alpha_j$s measure the change of the phase of the spinor part of 
the wavefunction as the system goes around a loop  $\Gamma_j$.  
The systems we are considering (see Fig.~\ref{fig:geometries})
are more simple in that the Schr\"odinger equation is separable 
if  the vector potential $\mathbf{A}(\mathbf{r})$ is chosen in an appropriate gauge, 
which takes into account the symmetry of the setup [i.e. translational
symmetry in the case of Fig.~\ref{fig:geometries}(a), (b) and
rotational in the case of Fig.~\ref{fig:geometries}(c)]. 
The magnetic field ${B}_z$ in Eq.~(\ref{eq:dtgamma}) will  
depend on only one of the (generalized) coordinates, let us   
denote this coordinate  by $x_1$, the conjugate momentum by $p_1$,
and the other coordinate (conjugate momentum) by $x_2$ ($p_2$). 
It turns out that due to the symmetry of the system  
$\gamma(\mathbf{r})$ will  also depend only on $x_1$\cite{ref:symmetries}. 
Therefore one of the two quantization conditions, involving the  coordinate 
$x_2$ and the conjugate momentum $p_2$, is exactly the same
as  it would be for a scalar wavefunction [this corresponds to  
$\alpha_2=0$ in Eq.~(\ref{eq:EBK_general})]. In the quantization
condition  involving  $p_1$ and $x_1$ however, the  phase $\alpha_1$ is  
in general not zero but is determined by Eq.~(\ref{eq:dtgamma}):
\begin{equation}
\alpha_1=\gamma_{\beta}^{\pm}=c_{\beta}^{\pm}\frac{\beta}{2}
\oint B_z(x_1(t)) dt.
\label{eq:alpha1}
\end{equation}
For systems with piece-wise constant magnetic field profiles such as those
shown in Fig.~\ref{fig:geometries}, the calculation of $\gamma_{\beta}^{\pm}$
simplifies to 
$
\gamma_{\beta}^{\pm}=c_{\beta}^{\pm}\frac{\beta}{2}\sum_{l}B_{z,l}T_l.
$
Here $T_l$ is the time the particle spends during one full
 period of its classical motion in the  $l$th region where 
the strength of the perpendicular component of 
the magnetic field is given by $B_{z,l}$. 
In the semiclassical picture $\gamma(\mathbf{r})$  changes only when 
the particle, during the course of
its classical motion, passes through  non-zero magnetic field regions 
and this phase change of the wave function needs to be taken 
into account in the semiclassical quantization.

\section{Bound states in graphene nanoribbons}
\label{sec:landaugauge}

We now apply the presented semiclassical formalism to determine the energy
of the bound states in inhomogeneous magnetic field setups
in graphene nanoribbons, see Figs.\ref{fig:geometries}(a) and (b).
Throughout the rest of the paper 
we will only consider $\mathcal{H}_{\beta}^{+}$  corresponding to positive energies.
$\mathcal{H}_{\beta}^{-}$ would describe
negative energies, these however do not need to be considered separately 
due to the chiral symmetry of the Hamiltonian, as explained in section \ref{sec:qm}. 

Using the Landau gauge  $\mathbf{A}=(0,A_y(x),0)^{T}$ 
the translation invariance of the system in the $y$ direction 
is preserved and therefore
the solution of the Hamilton-Jacobi equation Eq.~(\ref{eq:HamiltJacob})
can be sought as $S_{\beta}^{}(\mathbf{r})=S_{\beta}^{}(x)+p_y y$, where
$p_y=const$.  Since the classical motion in the $y$ direction is 
not bounded, $p_y=\hbar k_y$ is not quantized, it appears as a  
continuous parameter in our calculations. In contrast, the 
motion in the $x$ direction is bounded due to the $x$ dependent 
magnetic field $B_z(x)$. Therefore the quantization condition  reads
\begin{equation}
%$
\frac{1}{\hbar}\oint p_{\beta}(x) dx +\gamma_{\beta} =2\pi(n+1/2).
\label{eq:quant-ribbon}
%$
\end{equation}
(Note that  the Maslov index is $\mu=2$.) 
It is useful to introduce at this point  the 
following dimensionless
parameters: the width of the non-magnetic region $\tilde{w}=W/l_B$,
the guiding center coordinate $\tilde{X}=k_y l_B$, both in  units of
$l_B$ (which is defined  in Sec.~\ref{sec:qm}).
 Throughout this paper we will use  $\tilde{w}=2.2$.

We start our discussion with the magnetic waveguide configuration shown in Fig.~\ref{fig:geometries}(a) in graphene monolayer. 
Introducing the dimensionless energy $\tilde{E}_{ml}=E l_B/\hbar v_F$,
one finds that 
for $|\tilde{X}| < \tilde{E}_{ml}$ there is one turning point in each of 
the  left and right magnetic  regions.
A simple calculation  gives  $\gamma_1=\pi$  
and writing out explicitly the result of the action integral from  
Eq.~(\ref{eq:quant-ribbon}) it follows that   
\begin{equation}
4 K_{ml} \tilde{w} + \pi \tilde{E}_{ml}^2 =2 n \pi, \qquad n=1,2,\dots.
\label{eq:waveguide_mono}
\end{equation}
Here we have introduced the dimensionless wave number
$K_{ml}=\sqrt{\tilde{E}_{ml}^2-\tilde{X}^2}$ 
and note that the phase change of the wave function 
due to  $\gamma({x})$ 
cancelled the phase contribution coming from the Maslov index.
Furthermore,  if $\tilde{X} >\tilde{E}_{ml}$ ($\tilde{X} < -\tilde{E}_{ml}$) 
there are two turning points in the left (right) magnetic  regions. 
One finds that also for this case $\gamma_1=\pi$ which again cancels 
the contribution from the Maslov index, thus for 
$|\tilde{X}|> \tilde{E}_{ml}$ the  semiclassical quantization yields 
\begin{equation}
\tilde{E}_n=\sqrt{2 n}, \quad n=1,2,\dots
\label{eq:sc_quant_homB}
\end{equation}
i.e. the energies are independent of $\tilde{X}$ (and hence $k_y$).
This  is the same as the exact quantum and 
the semiclassical\cite{ref:ullmo} results for the relativistic Landau 
levels (LLs) in  \emph{homogeneous}
magnetic field. [From the exact quantum calculations\cite{ref:ghosh} it is known 
that  a zero energy state also exists in this system. 
Formally, from Eq.~(\ref{eq:sc_quant_homB}) one can 
obtain a zero-energy state by assuming that $n=0$ is admissible. 
However,  Eq.~(\ref{eq:V}) and hence Eq.~(\ref{eq:dtgamma}) are only valid
for $E\neq 0$.  Therefore we exclude $n=0$.]

\begin{figure}[htb]
\includegraphics[scale=0.42]{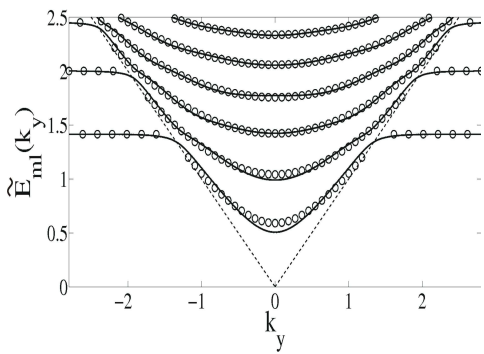}
\caption{Results of exact quantum calculations (solid lines) and 
the semiclassical approximation given by 
Eqs.~(\ref{eq:waveguide_mono}) and (\ref{eq:sc_quant_homB}) (circles) 
as a function of $k_y$ (in units of $l_B$) for graphene monolayer.
The dashed lines indicate $|\tilde{X}|=\tilde{E}_{ml}$ 
(see text). 
\label{fig:simcase-mono}}
\end{figure}

Comparison of the semiclassical eigenvalues with the results of 
exact quantum calculations are shown in Fig.~\ref{fig:simcase-mono}.
(For details of the quantum calculation see e.g. 
Ref.~\onlinecite{ref:ghosh}.)
The agreement between the quantum and semiclassical calculations is in
general very good, especially for higher energies. 
For lower energies and  $|\tilde{X}|\gg \tilde{E}_{ml}$
one can observe  quantum states  which are almost  
dispersionless and their energy is very close to  
the non-zero energy   LLs in graphene monolayer.
Semiclassically, these states are described by Eq.~(\ref{eq:sc_quant_homB}).
Although the zero energy state of the spectrum\cite{ref:ghosh} 
 can not be accounted for by our semiclassics
 an expression for  the gap  between the zero  and the first 
non-zero energy states 
can be easily obtained by putting  $n=1$ and $\tilde{X}=0$ in
Eq.~(\ref{eq:waveguide_mono}) and it gives a rather accurate prediction as
it can be seen in Fig.~\ref{fig:simcase-mono}. 
The presented semiclassical method can not describe those quantum states
which correspond to $|\tilde{X}| \approx \tilde{E}_{ml}$ 
[see the dashed line in Fig.~\ref{fig:simcase-mono}] i.e. when one of the 
the turning points is in the area of rapid spatial variation of the 
magnetic field.

For comparison, we have also calculated the quantization condition 
for graphene bilayer using the classical Hamiltonian given in
Eq.~(\ref{eq:Ham_bilay_Delta}) and the general quantization condition shown in 
Eq.~(\ref{eq:quant-ribbon}). For $\tilde{E}_{bl}>\tilde{X}^2/2$, where 
 $\tilde{E}_{bl}=\frac{ E\eta}{\hbar \omega_c}$, $\omega_c=\frac{|eB|}{m}$ being the 
cyclotron frequency, it reads:
\begin{equation}
%4 \sqrt{K_b^2-k_y^2}\, w + \pi (K_b l_B)^2 =2\pi\left(n-\frac{1}{2}\right)
 2 K_{bl}\, \tilde{w} + \pi \tilde{E}_{bl}  =\pi\left(n-\frac{1}{2}\right), 
\qquad n=2,3,\dots.
\label{eq:waveguide_bi}
\end{equation}
Here $K_{bl}=\sqrt{2 \tilde{E}_{bl}-\tilde{X}^2}$ 
and we have taken into account that in this case $\gamma_2=2\pi$.
For $\eta=1$ [where $\eta$ has been defined after Eq.~(\ref{eq:Ham_bilay_Delta})] 
this result is very similar to what one would
obtain for a 2DEG, the only difference
being that for 2DEG one would have $+1/2$ on the right hand side of 
Eq.~(\ref{eq:waveguide_bi}). This  similarity is a consequence 
of having a parabolic dispersion relation $E(\mathbf{k})$ 
close to the Fermi energy in both a 2DEG and graphene bilayer systems. 
\begin{figure}[htb]
\includegraphics[scale=0.42]{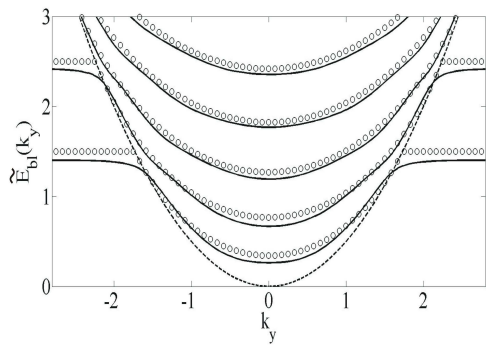}
\caption{Results of exact quantum calculations (solid lines) and 
the semiclassical approximation given by 
Eqs.~(\ref{eq:waveguide_bi}) and (\ref{eq:sc_bil_homB}) (circles) 
as a function of  $k_y$  (in units of $l_B$) for graphene bilayer.
The dashed lines indicate $\tilde{X}^2/2=\tilde{E}_{bl}$ (see text).
\label{fig:simcase-bi}}
\end{figure}
We let the integer quantum number $n$ to run from $n=2$  
in Eq.~(\ref{eq:waveguide_bi}) for the 
following considerations: from 
Ref.~\onlinecite{ref:ullmo} we know that in a more simple case of 
homogeneous magnetic field the 
four-fold degenerate\cite{ref:bilayer_degeneracy} LL
%Landau level 
of the quantum calculations\cite{ref:falko} 
at $\tilde{E}_{bl}=0$ (corresponding to $n=0,1$) 
can not be correctly described semiclassicaly but  
for LLs having $ \tilde{E}_{bl}>0 $ the agreement between the semiclassical
and quantum results is qualitatively very good.
Similarly, we expect that in our case the semiclassical approximation
should only work for $n\ge 2$. We have found that this is indeed
the case, see Fig.~\ref{fig:simcase-bi}  where the solid lines show bands 
 obtained by TB calculations and the circles are calculated using  
 Eq.~(\ref{eq:waveguide_bi}) for $n\ge 2$ [we have taken $\eta=1$].
The $\tilde{E}_{bl}>0$ energy bands for $\tilde{E}_{bl}>\tilde{X}^2/2 $  
are remarkably  well 
described by Eq.~(\ref{eq:waveguide_bi}) 
[note however that like in the homogeneous magnetic field case, 
there is a four-fold degenerate state at $\tilde{E}_{bl}=0$ ].  
For $\tilde{E}_{bl} < \tilde{X}^2/2 $ 
the bands of TB calculations again become almost dispersionless and level 
off  very close to the LLs %Landau levels
 of bilayer graphene in homogeneous field\cite{ref:falko}. 
The semiclassical 
expression for the energy levels in this regime of $\tilde{X}$ is 
\begin{equation}
\tilde{E}_{bl}=(n-1/2),\qquad n=2,3,\dots
\label{eq:sc_bil_homB}
\end{equation} 
which is again a good approximation of the quantum result. 
Our semiclassics can not correctly account for states having $\tilde{X}^2/2\approx\tilde{E}_{bl}$ i.e.
when one of the turning points is in rapidly varying magnetic field region.

We now turn to  the semiclassical study  of the system depicted in  Fig.~\ref{fig:geometries}(b), where 
the magnetic field is reversed in one of the regions. 
It has been shown 
in Refs.~\onlinecite{ref:rakyta,ref:ghosh} that   peculiar type of current 
carrying quantum states called snake states exist close to the 
$\mathbf{K}$ point of graphene for this magnetic 
field configuration. 
These states can also be described by the Dirac Hamiltonian and are therefore 
amenable to semiclassical treatment. 

 We start the discussion with the graphene monolayer  case. 
There are no turning points and hence no states if $-\tilde{X}>\tilde{E}_{ml}$.
For  $|\tilde{X}|< \tilde{E}_{ml}$ there is one turning point in each 
of the non-zero magnetic field regions.
In contrast to the symmetric magnetic field configuration 
[Fig.~\ref{fig:geometries}(a)]
we find  that for one full period of motion $\gamma_1=0$: 
the contributions of the two magnetic field regions, pointing in the 
opposite direction, cancel. Using Eq.~(\ref{eq:quant-ribbon}) 
the quantization condition is: 
\begin{equation}
\begin{split}
K_{ml} (2\tilde{w}+ \tilde{X})+
\tilde{E}_{ml}^2\left(\arcsin\left[\frac{\tilde{X}}{\tilde{E}_{ml}}\right]
+\frac{\pi}{2} \right)&=\pi(n+1/2) \\
n&=0,1,2,\dots
\end{split}
\label{eq:snake_mono}
\end{equation} 
Note, that unlike in the case of Eq.~(\ref{eq:waveguide_mono}) 
here  a solution for $n=0$ also exists. Moreover,
for $X>\tilde{E}_{ml}$ there are two turning points  both in the left 
(we denote them by $x_1^{L}$, $x_2^{L}$) and 
in the right (denoted by $x_1^{R}$, $x_2^{R}$) magnetic region. 
The  quantization using $x_1^{L}$, $x_2^{L}$ leads  to the 
same result as in Eq.~(\ref{eq:sc_quant_homB}) i.e. $\tilde{E}_{n,L}=\sqrt{2n}$,  
while using  $x_1^{R}$, $x_2^{R}$  gives the sequence $\tilde{E}_{n,R}=\sqrt{2(n+1)}$. 
The difference between  $\tilde{E}_{n,L}$ and $\tilde{E}_{n,R}$  
 is due to the fact that 
the sign of the phase contribution from $\gamma({x})$ depends on the 
direction of the magnetic field, i.e. it is $+\pi$ when $x_1^{L}$, $x_2^{L}$ is 
used in the calculations and $-\pi$ when $x_1^{R}$, $x_2^{R}$ is used [see Eq.~(\ref{eq:dtgamma})].  
From these considerations
it follows  that if $n=0,1,2,\dots$ as we assumed 
in Eq.~(\ref{eq:snake_mono}), 
the two sequence $\tilde{E}_{n,L}$, $\tilde{E}_{n,R}$  
 give two-fold degenerate dispersionless states  
at $\tilde{E}_{ml}=\sqrt{2}, \sqrt{4},\sqrt{6}\dots$ 
and a nondegenerate one at  $\tilde{E}_{ml}=0$. 
We have to exclude, however, the  $\tilde{E}_{ml}=0$ solution, see 
the discussion below Eq.~(\ref{eq:sc_quant_homB}).

\begin{figure}[htb]
\includegraphics[scale=0.42]{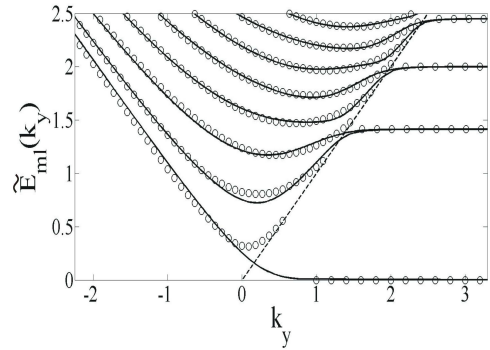}
\caption{Results of exact quantum calculations (solid lines) and 
the semiclassical approximation given by 
Eq.~(\ref{eq:snake_mono}) (circles) as a function $k_y$ (in units of $l_B$) 
for graphene monolayer.
The dashed line indicate $\tilde{X}=\tilde{E}_{ml}$. (see text).
\label{fig:snakecase-mono}}
\end{figure}

The results  of  quantum and of  semiclassical 
 calculations are shown in Fig.~\ref{fig:snakecase-mono}.
(For details of the quantum calculation see e.g. Ref.~\onlinecite{ref:rakyta}.)
As one can see the agreement is again very good for $\tilde{E}_{ml} \gtrsim 0.4$ 
except when $\tilde{X}\approx \tilde{E}_{ml}$, see the discussion 
in the previous example, i.e. when the magnetic fields point in the same
direction.  The two-fold degenerate 
dispersionless quantum states can also be observed for $\tilde{X}>\tilde{E}_{ml}$.
States having energies  $0 <\tilde{E}_{ml} \lesssim 0.25$ 
can not be described by our semiclassics and note that 
in the lowest energy band corresponding to $n=0$  
in Eq.~(\ref{eq:snake_mono}) nonphysical solutions also appear 
along with the genuine ones for $0.25 \lesssim \tilde{E}_{ml} \lesssim 0.4$. 
This clearly indicates the limits of applicability of our approach, 
i.e. it does not work for energies close to the Dirac point. 

We end our discussion of the bound states in mono and bilayer 
graphene nanoribbons with the  bilayer system corresponding to the 
previous, monolayer example, e.g. for the magnetic field
setup of Fig.~\ref{fig:geometries}(b).  
For  $-\tilde{X} > \sqrt{2\tilde{E}_{bl}}$ 
there are no turning points and hence no states. 
The quantization condition for  $\tilde{X}^2/2 < \tilde{E}_{bl}$ can simply
be obtained from  Eq.~(\ref{eq:snake_mono}) by  changing 
$\tilde{E}_{ml}\rightarrow \sqrt{2\tilde{E}_{bl}}$ and
$K_{ml}\rightarrow K_{bl}$
[$K_{bl}$ is defined after Eq.~(\ref{eq:waveguide_bi})].
\begin{figure}[htb]
\includegraphics[scale=0.42]{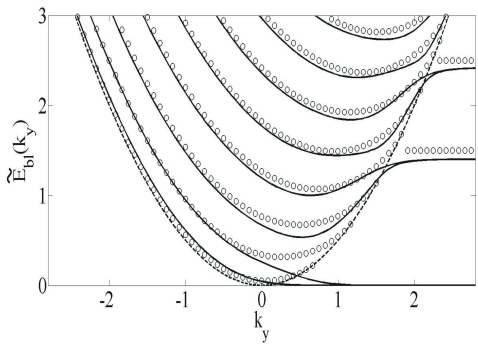}
\caption{
% Results of TB calculations (solid lines) and 
% the semiclassical approximation that can be obtained from 
% Eq.~(\ref{eq:snake_mono}) (circles) for graphene bilayer 
% as a function of $k_y$ (in units of $l_B$). 
% The dashed line indicate $\tilde{X}^2/2=\tilde{E}_{bl}$ (see text).
Results of TB calculations (solid lines) and  a  semiclassical approximation 
(circles) for graphene bilayer  as a function of $k_y$ (in units of $l_B$). 
The semiclassical approximation can be obtained from  Eq.~(\ref{eq:snake_mono}) 
by a transformation described in the main text.
% as a function of $k_y$ (in units of $l_B$). 
 The dashed line indicate $\tilde{X}^2/2=\tilde{E}_{bl}$ (see text).
\label{fig:snakecase-bi}}
\end{figure}
Finally,  for $\tilde{X}^2/2>\tilde{E}_{bl}$ our semiclassics predicts  
a sequence of doubly-degenerate, dispersionless  energy levels at $\tilde{E}_{bl}=(n-1/2)$, $n=2,3,\dots$, in a similar fashion as  
in the monolayer case.
As one can see in Fig.~\ref{fig:snakecase-bi} for $\tilde{E}_{bl} \gtrsim 0.8$
the semiclassical approximation captures all the main features of the TB 
calculations quite well, apart from the region where 
$\tilde{X}^2/2\approx\tilde{E}_{bl}$ for $\tilde{X}>0$ 
[see the discussion below Eq.~(\ref{eq:sc_bil_homB})].  
Dispersive states corresponding to $n=0,1$ can also be described 
semiclassically if $\tilde{E}_{bl} \gtrsim 0.8$ [see the lowest two bands in Fig.\ref{fig:snakecase-bi}],
but for smaller energies nonphysical solutions along with the genuine ones
do appear and for $\tilde{E}_{bl}\lesssim 0.1$ i.e. very close to the
Dirac point no quantum states can be described with the presented semiclassical
approach.

\section{Bound states of a magnetic quantum dot}
\label{sec:symmgauge}

Our last  example is the   magnetic dot in graphene monolayer discussed in 
Ref.~\onlinecite{ref:martino} and shown in Fig.~\ref{fig:geometries}(c).
We assume that the magnetic field is zero in  a circular region of radius $R$ 
while outside this region a  constant perpendicular  field is applied.

Working in polar coordinates $r$ and $\vfi$, 
since the vector potential of Eq.~(\ref{eq:vect_pot_pol}) preserves
the circular symmetry of the problem, the $\vfi$ coordinate is cyclic and 
therefore one can seek the solution of 
Eq.~(\ref{eq:HamiltJacob}) as $S(\mathbf{r})=S_r(r) + p_{\varphi}\varphi$ where 
$\pfi=const$. Moreover, in $\vfi$ coordinate the motion is free rotation 
and hence the quantization condition for $\pfi$ is simply 
\begin{equation}
\int_{0}^{2\pi}\pfi d\vfi=2\hbar\pi m, \qquad m=0,\pm 1,\pm 2\dots
\label{eq:pfi_quant}
\end{equation}
 whence it is clear that the $\pfi$  quantization reads  
$\pfi=\hbar m$.%\cite{ref:pfi}. 

The second quantization condition can generally be written as 
$
\frac{1}{\hbar}\oint p_r(r,m)dr +\gamma_1=2\pi(n+1/2)
$
because the Maslov index is $\mu=2$ and  $p_r=\frac{\prt S_r(r)}{\prt r}$. 
We now introduce the dimensionless variable $\xi = \frac{r^2}{2 l_B^2}$ and 
the parameters $\delta = \frac{R^2}{2 l_B^2}=\tilde{R}^2/2$, 
$\tilde{m} = m - \delta$.
One can see that $\delta$ is basically the 
missing magnetic flux that can be associated with the dot.
The  phase accumulated between  two points $\xi_1$, $\xi_2$
 inside the dot (where $B_z(r)=0$) 
is given by 
\begin{equation}
S_r^{B=0}(\xi_1,\xi_2) = \frac{\hbar}{2}\int_{\xi_1}^{\xi_2} 
{\rm d}\xi\frac{\sqrt{2\tilde{E}^2\xi - m^2}}{\xi}
\label{eq:s_r_B0}
\end{equation}
while between two points in the non-zero 
magnetic field region by 
\begin{equation}
S_r^{B\neq 0}(\xi_1,\xi_2) = \frac{\hbar}{2}\int_{\xi_1}^{\xi_2} 
{\rm d}\xi\frac{\sqrt{-\xi^2 + 2(\tilde{E}_{}^2-\tilde{m})\xi - \tilde{m}^2}}{\xi}
\label{eq:s_r_B}
\end{equation}
[the $\hbar$ factor in the above expressions appears because  
in the  calculation  we  already took into account the quantization
of $\pfi$, see Eq.~(\ref{eq:pfi_quant})].  The integrals 
in Eqs.~(\ref{eq:s_r_B0}) and (\ref{eq:s_r_B}) can 
be analytically calculated but the resulting expressions are too lengthy to be
recorded here. 

\begin{figure}[htb]
\includegraphics[scale=0.42]{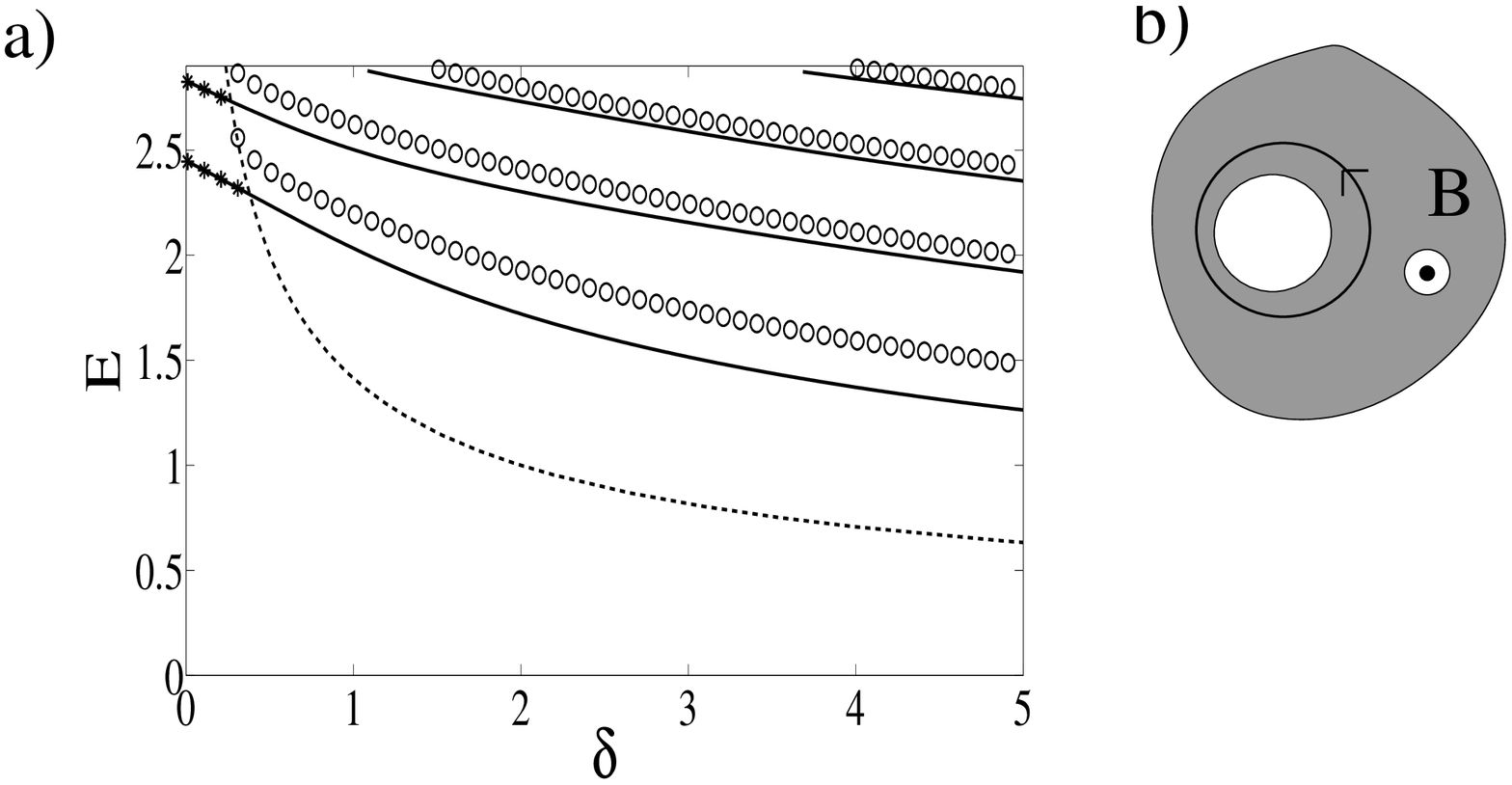}
\caption{(a) results of exact quantum calculations (solid lines) 
   for the $m=2$ energy bands as a function of the missing flux $\delta$.
   The results of the semiclassical quantization ($\ast$) obtained 
   from Eq.~(\ref{eq:sc_quant-1}) for 
   $\tilde{E}<\frac{|m|}{\sqrt{2\delta }}$, $\tilde{m}>0$. %are indicated by $\ast$. 
   For $\tilde{E}> \frac{|m|}{\sqrt{2\delta }}$ circles ($\circ$) show the 
   semiclassical results calculated using 
   Eqs.~(\ref{eq:sc_quant-2}) and (\ref{eq:gamma-2}).
   %for $\tilde{E}> \frac{|m|}{\sqrt{2\delta }}$. 
   The dashed line shows the $\tilde{E}=\frac{|m|}{\sqrt{2\delta}}$ function
   which separates the cases i) and ii) detailed in the main text. (b) shows
   a cartoon of a classical orbit in the parameter 
  range $\tilde{E}<\frac{|m|}{\sqrt{2\delta }}$, $\tilde{m}>0$.
\label{fig:sc_pol_m_2}}
\end{figure}

As a next step to obtain a semiclassical quantization rule 
we proceed with the analysis of the classical dynamics along the 
lines of Ref.~\onlinecite{ref:kocsis}. 
Calculating the  radial velocity 
$v_r(r)=\frac{\hbar v_F}{E} p_r$
in the magnetic region  one finds that  
for $\tilde{E}^2_{}>2\tilde{m}$ there can be two  
turning points in the radial motion which we denote 
by $\xi_0^-$, $\xi_0^+$.  In terms of the dimensionless parameters
 $\tilde{R}_c=\tilde{E}$ and $\tilde{X}=\sqrt{\tilde{R}_c^2-2\tilde{m}}$
(the radius of the 
classical cyclotron motion and the  guiding center coordinate  
respectively, both in the units of $l_B$) 
the turning points can be written as  
$
\xi^{\pm}_0=\frac{1}{2}(\tilde{R}_c\pm \tilde{X} )^2.
$
With regard to $\delta$, there are then two possible cases.

i) The first case is when $\delta <\xi_0^-, \xi_0^+$ (or equivalently, $\delta<\frac{m^2}{2\tilde{E}^2}$) and therefore 
 the radial motion is 
   confined entirely to the magnetic region. 
   Calculation of 
   $\gamma_1$ gives a phase $+ \pi$ which cancels the phase 
    contribution from the Maslov index. Hence 
   the quantization condition is  
   $
   2 S_r^{B\neq 0}(\xi_0^-,\xi_0^+)/\hbar=2n\pi
   $
   where $ S_r^{B\neq 0}(\xi_0^-,\xi_0^+)$ is calculated using 
   Eq.~(\ref{eq:s_r_B}).
   Explicitly, the energy levels $\tilde{E}_{n,m}$ are given by 
\begin{equation}
\tilde{E}_{n,m} = \sqrt{2n + \left|\tilde{m}\right| + \tilde{m}}.
%        \quad\mbox{if}\quad\delta<\frac{m^2}{2\tilde{E}^2}
\label{eq:sc_quant-1}
\end{equation}
This result is very similar to what one would 
obtain from exact quantum calculations for a homogeneous 
magnetic field, where the relativistic Landau levels are given by 
$
 \tilde{E}_{n,m}^{qm} = \sqrt{2n + \left|{m}\right| + {m}}.
$
\begin{figure}[htb]
\includegraphics[scale=0.42]{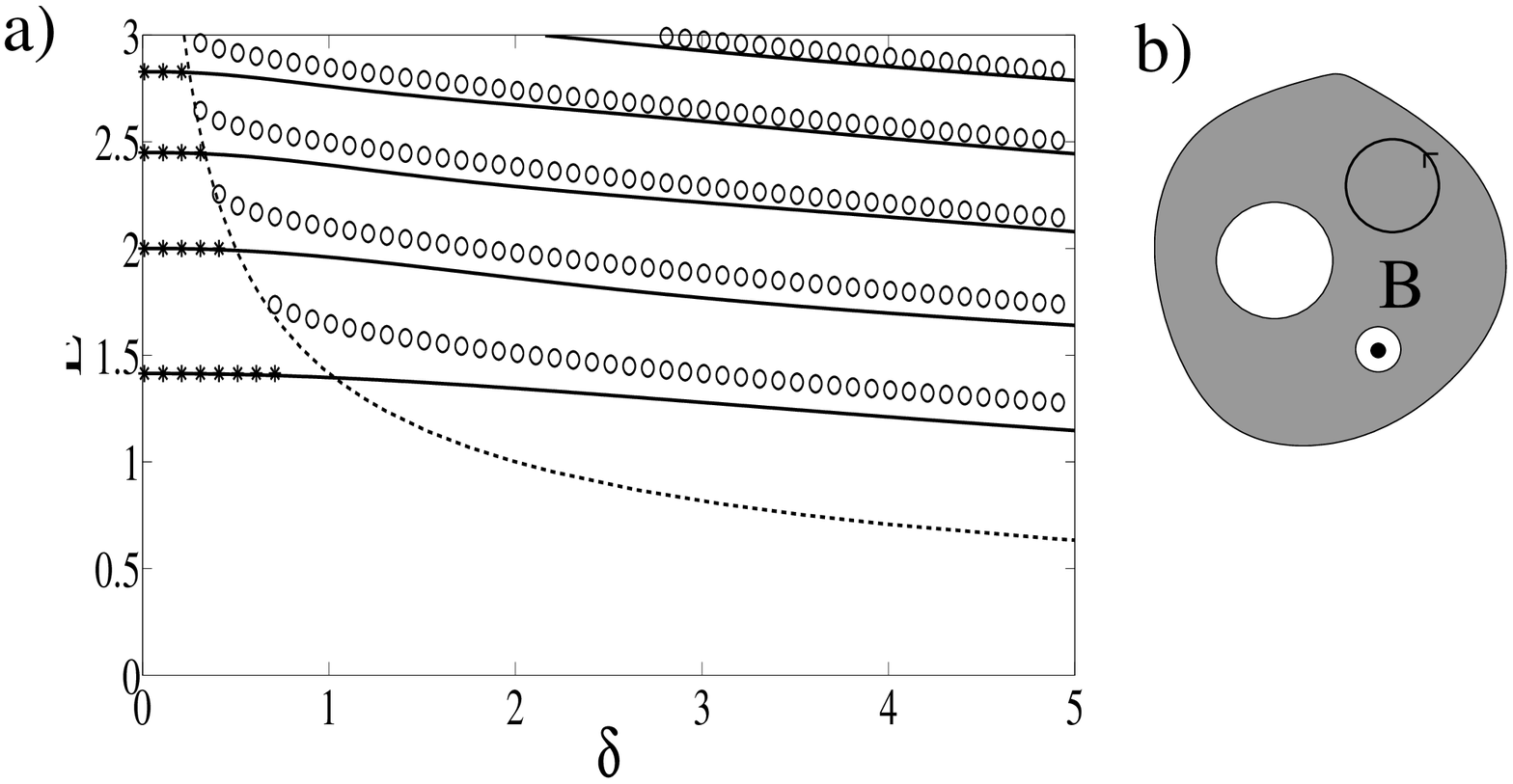}
\caption{a) results of exact quantum calculations (solid lines) 
   for the $m=-2$ energy bands as a function of the missing flux $\delta$.
   The results of the semiclassical quantization ($\ast$) obtained 
   from Eq.~(\ref{eq:sc_quant-1}) for 
   $\tilde{E}<\frac{|m|}{\sqrt{2\delta }}$, $\tilde{m}<0$. %are indicated by $\ast$. 
   For $\tilde{E}> \frac{|m|}{\sqrt{2\delta }}$ circles ($\circ$) show the 
   semiclassical results calculated using 
   Eqs.~(\ref{eq:sc_quant-2}) and (\ref{eq:gamma-2}).
   The dashed line shows the $\tilde{E}=\frac{|m|}{\sqrt{2\delta}}$ function
   which separates the cases i) and ii) detailed in the main text. (b) shows
   a cartoon of a classical orbit in the parameter 
  range $\tilde{E}<\frac{|m|}{\sqrt{2\delta }}$, $\tilde{m}<0$.
\label{fig:sc_pol_m_-2}}
\end{figure}
Note however that in  Eq.~(\ref{eq:sc_quant-1}) instead of the integer 
quantum number $m$ the non-integer  $\tilde{m}=m-\delta$ appears.
From Eq.~(\ref{eq:sc_quant-1}) it is clear that for $\tilde{m}<0$   
the energy bands are at 
$\tilde{E}_n=\sqrt{2 n}$, $n=1,2,\dots$  and they do not depend on $\delta$ and $\tilde{m}$. (As in the nanoribbon case, we exclude $n=0$ because that would give $\tilde{E}_n=0$.) 
These $\delta$ independent sections are readily observable in the energy 
bands corresponding to $m=-2$ in Fig.~\ref{fig:sc_pol_m_-2}(a).
On the other hand, for  $\tilde{m} >0$,  $\delta \ll m$   
an approximately linear dependence on $\delta$ of 
the bands corresponding to different $m$ is predicted 
by  Eq.~(\ref{eq:sc_quant-1}) and this  can also be observed, see  Fig.~\ref{fig:sc_pol_m_2}(a).
From classical point of view in the parameter range 
 $\tilde{m}<0$  (which  
implies $\tilde{X}-\tilde{R}_c>\tilde{R}$) the classical orbits are such that they
 do not encircle the zero magnetic field region 
[see Fig.~\ref{fig:sc_pol_m_-2}(b)]. 
They are just 
like orbits in homogeneous magnetic field and this helps to understand
why the quantum states corresponding to the same parameter range 
are reminiscent of  dispersionless Landau levels. Conversely, 
for $\tilde{m}>0$ (which implies $\tilde{R}_c-\tilde{X}>\tilde{R}$)
the classical orbits do encircle the zero magnetic field region
and therefore the energy of the  corresponding quantum states depend
on the missing flux $\delta$ [see the cartoon shown in Fig.~\ref{fig:sc_pol_m_2}(b)
for illustration of the classical orbits].

ii) The second case is when $ \xi_0^-< \delta < \xi_0^+$ which
   happens  if $\delta>\frac{m^2}{2\tilde{E}^2}$. 
   The classical
   motion is no longer confined to the magnetic region but also
   enters the nonmagnetic dot. The turning point in the  nonmagnetic
    region is at $\xi_0=\frac{m^2}{2 \tilde{E}^2}<\delta$. 
    Since the Maslov index is $\mu=2$, the quantization condition can 
    be written as 
\begin{equation}
\frac{2}{\hbar}\left[S_r^{B=0}(\xi_0,\delta)+S_r^{B\neq 0}(\delta,\xi_0^+)\right]
+\gamma_1=2\pi(n+1/2).
\label{eq:sc_quant-2}
\end{equation} 
Here $S_r^{B=0}(\xi_0,\delta)$ and $S_r^{B\neq 0}(\delta,\xi_0^+)$
can be calculated using Eq.~(\ref{eq:s_r_B0}) and Eq.~(\ref{eq:s_r_B}) respectively,
 but the resulting expressions are again too lengthy to be presented here. 
Moreover, from Eq.~(\ref{eq:alpha1})  we find that 
\begin{equation}
\gamma_1 =  \frac{\pi}{2} + \arcsin\left(\frac{\tilde{E}^2 - \tilde{m}-\delta}{\sqrt{\tilde{E}^2(\tilde{E}^2-2\tilde{m})}}\right).
\label{eq:gamma-2}
\end{equation}
One can see that  here
in general $\gamma_1 \neq \pi$ and therefore it 
does not cancel the contribution of the Maslov index.

As one can observe the overall agreement of the exact quantum and 
of the semiclassical calculations  shown in 
Figs~\ref{fig:sc_pol_m_-2},~\ref{fig:sc_pol_m_2} is good, especially 
for higher energies. According to the exact 
quantum calculations\cite{ref:martino}, there is
also a zero energy state but this can not be described by our semiclassics.

\section{Conclusions}
\label{sec:concl}

To conclude, using semiclassical quantization 
we have studied the spectrum of bound states
in inhomogeneous magnetic field setups in graphene mono and bilayer. 
We have found that a semiclassical quantization which takes
into account a "Berry-like"  phase can indeed explain all the main features 
of the exact quantum or numerical TB calculations. In particular, we have 
studied graphene mono-and bilayer nanoribbons in magnetic 
waveguide configuration and also in a configuration when snake  
states can exist. Besides, we discussed the magnetic dot system
in graphene monolayer.  For the considered step-wise constant 
magnetic field profile we have  derived semiclassical quantization equations. 
In the case of graphene monolayer, we have compared the resulting 
semiclassical eigenenergies to quantum mechanical ones obtained 
from the corresponding Dirac equation. For graphene bilayer 
the results of the semiclassical quantization and
numerical TB calculations have been compared.
In all the cases a  good agreement
has been found except for energies very close to the Dirac point.  
We have shown that the main features of the spectrum depend on
whether the classical guiding center coordinate is in the 
non-magnetic or in the magnetic field region.

Assuming homogeneous magnetic field, the energy of the 
Landau levels in semiclassical approximation 
has been calculated in Ref.~\onlinecite{ref:ullmo}. 
Our work can be considered as a generalization of these  
calculations to a class of non-homogeneous magnetic field setups,
where due to the symmetry of the system,
the Berry-like phase appearing in the semiclassical theory 
affects only one of the quantization conditions.

\section{Acknowledgment}
We acknowledge discussions with Henning Schomerus. 
This work is supported partly by European Commission Contract
No.~MRTN-CT-2003-504574 and by EPSRC.
J. Cs. would like to acknowledge the support of the 
Hungarian Science Foundation OTKA  T48782.


\begin{thebibliography}{10}
\bibitem{ref:bilayer-tb}
To our best knowledge tight-binding (TB) or analytic calculations for
the systems shown in Fig.~\ref{fig:geometries}(a),(b) 
have not yet been published for bilayer graphene.   
We obtained the results presented in Sec.\ref{sec:landaugauge} 
using the   TB Hamiltonian given  e.g. in Ref.~\onlinecite{ref:e-castro}.
In  the numerical calculations the  width of the  
graphene ribbon was $L=(\sqrt{3} N -1)a$ where 
$N=2400$ is the number of lattice sites in the 
$x$ direction. Furthermore, we used $t_2/t_1=0.156$ and
the strength of the magnetic field was $1T$.
\bibitem{ref:bilayer_degeneracy}
For the magnetic field used in our TB calculations the sequence of 
the Landau levels in bilayer is given by $E_n=\hbar\omega_c\sqrt{n(n-1)}$ 
see Ref.~\onlinecite{ref:falko}. Therefore the $n=0$ and $n=1$ levels are
degenerate and another two-fold degeneracy comes from the electron-hole
symmetry of the Hamiltonian.
\bibitem{ref:symmetries}
Generally, using  group theoretical arguments one can show   
that  if the system has a continous symmetry, described by a one parameter
symmetry transformation $\hat{S}(\theta)$, then 
from Eq.~(\ref{eq:dtgamma}) it follows that $\gamma(\mathbf{r})$ 
 can only depend on the fundamental invariant $F(\mathbf{r})$ of 
 $\hat{S}(\theta)$. Here $F(\mathbf{r})$ is a particular function of the 
 coordinates and together with an appropriately chosen $G(\mathbf{r})$ 
 function they can define a new (curvilinear) coordinate system. In this
 new coordinate system then our considerations  leading to Eq.~(\ref{eq:alpha1}) 
do apply. For the graphene monolayer examples, where the  exact quantum 
 solutions of the problems are known, this can also be  verified by taking 
the short-wavelength assymptotic form of the  quantum wavefunctions.
\end{thebibliography}
\end{document}